\documentstyle[12pt,prl,aps]{revtex} 
\large
\def\addcontentsline#1#2#3{\relax}

\long\outer\def\demo#1. #2\par{\medbreak\noindent {\bf#1.\enspace}
        {\rm#2}\par\ifdim\lastskip<\medskipamount\removelastskip
        \penalty55\medskip\fi}
\newcommand{\ben}{\begin{enumerate}}
\newcommand{\een}{\end{enumerate}}
\def\bdemo#1. #2\par{\medbreak\noindent {\bf#1.\enspace}{\rm#2}\par}
\def\edemo{\ifdim\lastskip<\medskipamount\removelastskip\penalty55\medskip\fi}

\def\0{{\bf 0}}
\def\1{{\bf 1}}

\def\c{{\bf c}}
\def\n{{\bf n}}
\def\x{{\bf x}}
\def\N{{\bf N}}
\def\Q{{\bf Q}}
\def\R{{\bf R}}
\def\D{{\rm D}}
\def\Exp{{\rm Exp}}
\def\diag{{\rm diag}}
\def\BD{\mathop{\rm BD}\nolimits}
\def\mod{\mathop{\rm mod}\nolimits}

\def\longto{\mathop{\longrightarrow}\limits}
\def\bfa{\mathop{a\kern-.5em{a}\kern-.5em{a}}\nolimits}
\def\bfb{\mathop{b\kern-.47em{b}\kern-.47em{b}}\nolimits}
\def\bfx{\mathop{x\kern-.6em{x}\kern-.6em{x}}\nolimits}
\begin{document}
\baselineskip 18pt
\author{Yshai Avishai$^{\,1,3}$, Daniel Berend$^{\,2}$
and Richard Berkovits$^{\,3}$}
\address{${}^{1}$Department of Physics, Ben-Gurion University 
of the Negev, Beer-Sheva 84 105, Israel \\
${}^{2}$Department of Mathematics and Computer Science,
 Ben-Gurion University of the Negev,
Beer-Sheva 84 105, Israel \\
${}^{3}$Minerva Center, Department of Physics, Bar-Ilan University,
Ramat Gan, Israel  }

\title{Statistics of addition spectra of independent quantum systems}

\maketitle

\begin{abstract}
 \\
Motivated by recent experiments on large quantum dots,
we consider the energy spectrum in a system consisting
of $N$ particles distributed among $K<N$ 
independent sub-systems, such that the energy of each 
sub-system is a quadratic function of the number of 
particles residing on it. On a large scale, the ground state
energy $E(N)$ of such a system grows 
quadratically with $N$, but in general there is no simple 
relation such as $E(N)=a N +b N^{2}$. 
The deviation of $E(N)$ from exact quadratic behavior implies that its second 
difference (the inverse compressibility) $\chi_{N} \equiv E(N+1)-2 E(N) 
+E(N-1)$ is a fluctuating quantity.
Regarding the numbers $\chi_{N}$ as values assumed by a certain random
variable $\chi$, we obtain a closed-form expression for 
its distribution $F(\chi)$. Its main feature is 
that the corresponding density $P(\chi)=\frac{dF(\chi)} {d\chi}$ 
has a maximum at the point $\chi=0$. As $K \rightarrow \infty$ 
the density is Poissonian, namely, $P(\chi) \rightarrow e^{-\chi}$.\\
\end{abstract}


\newpage
\section {Motivation}
Statistics of spectra is an efficient tool for elucidating 
properties  of various physical systems. So far, most of 
the effort is focused on the study of energy levels of a system
with a fixed number of particles. In this 
context, one of the central earlier results 
is that the spectral statistics of many-body 
systems such as complex nuclei agree with the predictions of 
random matrix theory \cite{Wigner,Dyson}. On the 
other extreme, it was found that level statistics of 
a single particle in chaotic or disordered system also 
obeys a Wigner-Dyson statistics \cite{Bohigas,Altshuler}.

Recently, experiments are designed to get information on 
the statistics of the {\em addition spectra} of electrons 
in quantum dots \cite{Sivan}. The pertinent energy levels
$E(N)$ are the ground state energies of a system consisting of $N$ electrons 
residing on a quantum dot, which is coupled capacitively to its environment.

Let us single out two properties of the addition 
spectra of quantum dots. The first one is that, on a large scale, 
the energy $E(N)$ grows quadratically with $N$, while the second one is 
a consequence of charge quantization, namely, there is, in general, no simple 
relation such as $E(N)=a N +b N^{2}$. In this context, an appropriate quantity 
whose statistics is of interest is then the inverse compressibility,
\begin{eqnarray}
\chi_{N} \equiv E(N+1)-2 E(N)+E(N-1).
\label{eq_compress}
\end{eqnarray}
It is the deviation of $E(N)$ from exact quadratic behavior which makes its
second difference $\chi_{N}$ non-constant. Indeed,
in a recent experiment on large quantum dot \cite{ZAPW} it was found that the
inverse compressibility vanishes for numerous values of electron number~$N$.

In the present work we study the statistics of the addition spectrum of a
simple physical system with the two basic properties mentioned above. One 
example of such a system is motivated by considering the electrostatic energy
of large quantum dots (although it should be mentioned that the model is too
simple to describe the actual physics).
To be specific, we have in mind a system of $K$ metallic 
grains such that the number of electrons on the $i^{th}$
grain is $n_{i}$ ($i=0,1,2,\ldots,K-1$) and their sum equals $N$. The
electrostatic energy of the pertinent system is a bilinear form in the 
numbers $n_{i}$ with a $K \times K$ matrix $w \equiv \frac {1} {2} C^{-1}$.
Here $C$ is a positive-definite symmetric matrix of 
capacitance and inductance coefficients.
If the metallic grains are very far apart, the matrix $C$ is nearly diagonal.
Thus, we concentrate on the special case $C=\diag[C_{i}]$, for which the energy
of the system is given by 
\begin{eqnarray}
E(N)=\min \sum_{i=0}^{K-1} \frac {1} {2 C_{i}}
n_{i}^{2}, \mbox {(subject to $\sum_{i=0}^{K-1} n_{i}=N).$}
\label{eq_EN}
\end{eqnarray} 
The minimum in (\ref{eq_EN}) is taken over all possible partitions $n_i$ of $N$.

Another example is the energy of a system composed of $K$ different harmonic
oscillators, among which one distributes $N$ spinless fermions.
If there are $n_{i}$ fermions on 
oscillator $i$ (whose frequency is $\omega_{i}$),
then the energy of this oscillator (up to a constant) is
$E_{i}=\hbar \omega_{i} n_{i} (n_{i}+1)$,
and hence the ground state energy of the system is
\begin{eqnarray}
E(N)=\min \sum_{i=0}^{K-1} E_{i},
\mbox {(subject to $\sum_{i=0}^{K-1} n_{i}=N).$}
\label{eq_ENOS}
\end{eqnarray}
We will concentrate on the first example, which is borrowed from
the electrostatics of quantum dots (\ref{eq_EN}), and 
refer to the constants $C_{i}$ as capacitors. Some results pertaining to the
second example (the system of oscillators (\ref{eq_ENOS})) are also presented.

Regarding the numbers $\chi_{N}$ of (\ref{eq_compress}) as values assumed by a 
certain random variable, the distribution of this random variable is the main
focus of the present work, which culminates in Theorem~1,
where we find a closed-form expression for the distribution.

The problem of elucidating the (addition) spectral statistics of a
{\em a many-body} system, consisting of several independent sub-systems 
(whose dependence of $E$ on $n_{i}$ is known), looks deceptively simple.
As will be evident shortly, this is not the case, and finding 
the distribution in question is quite a non-trivial task. Note that,
even for a {\em single particle system} composed of several 
independent sub-systems ({\em e.g.,} a system of a particle in several boxes),
the derivation of level statistics requires a large degree 
of mathematical effort \cite{Berry}. The rest of the paper is therefore devoted
to a rigorous derivation of our main results.
\section{Formalism}
\demo Definition 1. Let $(\theta_n)_{n=1}^\infty$ be a sequence of real numbers
and $F$ a distribution function. The sequence $(\theta_n)$ is {\it asymptotically
$F$-distributed} if
$$\frac{\left|\{1\le n\le M:\theta_n\le x\}\right|} {M}\longto_{M\to\infty} F(x)$$
for every continuity point $x$ of $F$ (where $|S|$ denotes the cardinality of
a finite set $S$).

An equivalent condition is the following. Denote by $\delta_t$ the point mass
at $t$, and let $\mu$ be the probability measure corresponding to the distribution
$F$ (namely, $\mu(A)=\int 1_A dF(x)$ for any Borel set $A$). Then $(\theta_n)$ is
asymptotically $F$-distributed if
$$\frac{1}{M}\left(\delta_{\theta_1}+\delta_{\theta_2}+\ldots+\delta_{\theta_M}\right)
\longto_{M\to\infty} \mu$$
(the convergence being in the weak*-topology).

The notion of asymptotic distribution has a stronger version whereby, instead of
requiring only that initial
pieces of the sequence behave in a certain way, we require this to
happen for any large finite portion of the sequence. This leads to

\demo Definition 2. In the setup of Definition 1, $(\theta_n)$ is
{\it asymptotically well $F$-distributed} if
$$\frac{\left|\{L< n\le M:\theta_n\le x\}\right|} {M-L}\longto_{M-L\to\infty} F(x)$$
for every continuity point $x$ of $F$.

Recall that the {\it density} of a set $A\subseteq\N$ is given by
$$\D(A)=\mathop{\lim}_{M\to\infty} {\left|A\cap [1,M]\right|\over M}$$
if the limits exists. If, moreover, the limit
$$\BD(A)=\mathop{\lim}_{M-L\to\infty} {\left|A\cap (L,M]\right|\over M-L}$$
exists (in which case it is certainly the same as $\D(A)$), then it is called
the {\it Banach density} of $A$ (cf. \cite[p.72]{Fu}).

The following lemma is routine.

\proclaim Lemma 1. Let $(\theta_n)_{n=1}^\infty$ be a sequence of real numbers.
Suppose $\N=\bigcup_{j=1}^r A_j$, where the union is disjoint. Let
$(\theta^{(j)}_n)_{n=1}^\infty$ be the subsequence of $(\theta_n)$, consisting of
those elements $\theta_n$ with $n\in A_j,\ 1\le j\le r$.
\ben
\item
If each $(\theta^{(j)}_n)$ is asymptotically $F_j$-distributed for some
distribution functions $F_j,\ 1\le j\le r$, and $\D(A_j)=d_j,\ 1\le j\le r$, then
$(\theta_n)$ is asymptotically $F$-distributed, where $F=\sum_{j=1}^r d_j F_j$.
\item
If each $(\theta^{(j)}_n)$ is asymptotically well $F_j$-distributed and
$\BD(A_j)=d_j$, then $(\theta_n)$ is asymptotically well $F$-distributed.
\een

Obviously, a general sequence on the line does not have to be asymptotically
distributed according to some distribution function, but one would expect it of
sufficiently ``regular" bounded sequences. In our case, one might expect $\chi_{N}$
to be distributed according to some distribution function corresponding to a measure
centered at about $1/C$. However, this is not the case. In fact, the measure in
question is supported on a finite interval, and is a convex combination of an
absolutely continuous measure with decreasing density function on some interval
$[0,a]$ and the point mass $\delta_a$ at the right end $a$ of that interval.

We have defined $E(N)$ indirectly by means of the following

{\bf Problem 1}. For each non-negative integer $N$, find non-negative integers
$n_0,n_1,\ldots,n_{K-1}$, satisfying $n_0+n_1+\ldots+n_{K-1}=N$, for which
$\sum_{i=0}^{K-1} \frac{1}{2C_i}\cdot n_i^2$ is minimal.

It turns out that this problem is intimately related to a second optimization problem.
Put $w_i=\frac{1}{2C_i},\ 0\le i\le K-1$, and let $\Delta$ denote the set of all
positive odd multiples of the numbers $\frac{1}{2C_i}$:
$$\Delta=\{w_0,3w_0,5w_0,\ldots,w_1,3w_1,5w_1,\ldots,
           w_{K-1},3w_{K-1},5w_{K-1},\ldots\}\;.$$
Here we treat $\Delta$ as a multi-set, or a sequence, in the sense that if some elements
appear in this representation of $\Delta$ more than once (which occurs if some ratio
$w_i/w_j$ is a rational number with odd numerator and denominator), then we
consider $\Delta$ as having several copies of these numbers.

{\bf Problem 2}. For each non-negative integer $N$, minimize
$\sum_{m=1}^N \delta_m$, where $\delta_1,\delta_2,\ldots,\delta_N$ range over all distinct
$N$-tuples in $\Delta$.

Note that, if an element appears several times in $\Delta$, it is allowed to appear the
same number of times in the sum as well.

Let us demonstrate the equivalence of the two problems. Given the sum
$\sum_{i=0}^{K-1} w_i\cdot n_i^2$, we may use the equality
$w_i\cdot n_i^2=w_i+3w_i+5w_i+\ldots+(2n_i-1)w_i$
to see that any feasible value for the objective function of the first problem is a
feasible value for the objective function of the second problem as well. On the other
hand, solving Problem 2 is trivial. Namely, one minimizes the sum there simply by
taking the $N$ least elements of the set $\Delta$. In particular, for each $i$, the
multiples of $w_i$ present in the optimal solution will be all odd multiples
$w_i,3w_i,5w_i,\ldots$ up to some $(2n_i-1)w_i$. Thus, the optimal solution of
Problem 2 yields the optimal solution of Problem 1 also. We note in passing that
this discussion shows also that the minimum (for each of the problems) is obtained
at a unique point unless $\Delta$ contains multiple elements. (However, we shall
always refer to {\bf the} optimal solution, even when there may be several.)

A simple consequence of the above is

\proclaim Proposition 1. Let $\n=(n_i)_{i=0}^{K-1}$ be the optimal solution
of Problem 1 for some value of $N$. Then the optimal solution of Problem 1,
with $N+1$ instead of $N$, is $\n'=(n'_i)_{i=0}^{K-1}$, where $n'_j=n_j+1$ for
some $0\le j\le K-1$ and $n'_i=n_i$ for $i\ne j$.

\demo Remark. It is convenient to comment here on the effect a certain change
in the original problem would make. One may consider the energies $E_i$ to
be $w_i n_i(n_i+1)$ instead of $w_i n_i^2$. This would change $\Delta$ to be
the set of all even multiples of the $w_i$'s. Obviously, this would leave intact the
equivalence of Problems~1 and 2. One can check that this would have also no effect
on Theorems~1 and~2 {\it infra}.

To formulate our main result we need a few definitions and notations. Real numbers
$\theta_1,\theta_2,\ldots,\theta_r$ are independent over $\Q$ if, considered as
vectors in the vector space $\R$ over the field $\Q$, they are linearly independent.
Equivalently, this is the case if the equality $m_1\theta_1+m_2\theta_2+\ldots+
m_r\theta_r=0$ for integer $m_1,m_2,\ldots,m_r$ implies $m_1=m_2=\ldots=m_r=0$.
Considering the actual physical system (a collection of metallic grains),
it is reasonable to assume that the capacitors $C_{i}$ are random, 
so that generically they are independent over $\Q$. Without loss of generality we 
may rearrange the $K$ capacitors such
that $C_{0}=\max_{0\le i\le K-1} C_{i}$. It is also useful to 
divide all the capacitors by the largest one, so that the 
scaled capacitors $c_{i} \equiv C_{i}/C_{0}$ with
$1=c_{0}>c_{1},c_{2}\ldots,c_{K-1}$ are dimensionless.
Finally, set $s=c_0+c_1+\ldots+c_{K-1}$.

Now we formulate our main results.

\proclaim Theorem 1. Suppose $C_0,C_1,\ldots,C_{K-1}$ are independent over $\Q$.
Then the sequence $(\chi_N)_{N=1}^\infty$ is asymptotically $F$-distributed, where
the distribution $F$ is given by either of the following two representations:
\begin{eqnarray}
F(x)&=&\cases{
               0,&$\ x<0,$\cr
               \displaystyle{1-\frac{1}{s}\sum_{i=0}^{K-1} c_i\prod_{j=0\atop j\ne i}^{K-1}
                   \left(1-\frac{x}{2w_j}\right),}&$\ 0\le x<2w_0,$\cr
               1,&$\ 2w_0\le x,$\cr
              }\label{eq_P_1(x)}\\\cr\cr\cr
    &=&\cases{
               0,&$\ x<0,$\cr
               \displaystyle{1-\frac{1}{s}\sum_{S\subseteq\{1,\ldots,K-1\}}}
                  \left(|S|+1\right)\prod_{i\in S} c_i
                  \prod_{i\notin S} (1-c_i)
                  \cdot\left(1-\frac{x}{2w_0}\right)^{|S|},&$\ 0\le x<2w_0,$\cr
               1,&$\ 2w_0\le x.$\cr
              }
\label{eq_P_2(x)}
\end{eqnarray}

It is not immediately obvious from the formulas, but $F$ has one discontinuity, namely
at the point $2w_0$. The reason is that, as the elements of $\Delta$ are all odd
multiples of the $w_i$'s, and as $w_0$ is the smallest of the $w_i$'s, it happens
occasionally that there is no odd multiple of $w_1,\ldots,w_{K-1}$ between two
consecutive multiples of $w_0$. The size of the atom at $2w_0$ is
$\frac{1}{s}\cdot\prod_{i=1}^{K-1} (1-c_i)$. This is easily explained intuitively.
In fact, the
``density" of odd multiples of $w_i$ is $c_i$ times the same density for multiples of
$w_0$. Hence the ``probability" that an interval of the form $[(2n-1)w_0,(2n+1)w_0)$
does not contain an odd multiple of $w_i$ is $1-c_i$. Assuming that the ``events"
of containing different $w_i$'s are independent, we conclude that the proportion
of multiples of $w_0$ in $\Delta$ whose successors are also such is
$\prod_{i=1}^{K-1} (1-c_i)$. Since the proportion of multiples of $w_0$ in $\Delta$
is $\frac{1}{s}$, we arrive at the required expression for the size of the atom.

Now we would like to study the asymptotic of the distances between consecutive
elements of $\Delta$ as the number of capacitors grows. Obviously, as this happens,
the distances become smaller. More precisely, on the average we have
$\frac{1}{2w_j}$ odd multiples of each $w_j$ in each unit interval, and hence we have
there $\sum_{j=0}^{K-1}\frac{1}{2w_j}=\frac{s}{2w_0}$ elements of $\Delta$ altogether.
Hence the average distance between consecutive elements is $\frac{2w_0}{s}$. To
understand the asymptotic of the gaps, it makes sense therefore to normalize them
so as to have mean 1. Thus, we multiply the distances by $\frac{s}{2w_0}$, and ask
about the asymptotic behavior

\proclaim Theorem 2. Suppose the capacitances $C_0,C_1,\ldots$ are chosen uniformly
and independently in $[0,1]$. For each $K$, let $F_K$ denote the distribution
corresponding to the normalized gaps when taking into account the first $K$
capacitors only. Then, with probability 1, the distributions $F_K$ converge
to an $\Exp(1)$ distribution function.

\demo Remark. As will be seen in the proof, we actually use much less to prove
Theorem 2 than is required by the conditions of the theorem. Namely, we need
the capacitances $C_i$ to be linearly independent over $\Q$, and that they do
not form a fast diminishing sequence.

It is worthwhile mentioning that this type of ``Poissonian" asymptotic behavior 
of consecutive gaps is typical. For example, this is the case for uniformly selected
numbers in $[0,1]$, and is conjectured to be the case in other interesting cases
as well. (See, for example, \cite{KR} and \cite{RZ} and the references there.)

In the course of the proof, we shall make use of the notion of uniform distribution
modulo~1 and a few basic results relating to it. (The reader is referred to
Kuipers and Niederreiter \cite{KN} for more information.) A sequence
$(x_n)_{n=1}^\infty$ of real numbers is {\it uniformly distributed modulo}~1 if
$$\frac{\left|\{1\le n\le M:a\le \{x_n\}<b\}\right|}{M}\longto_{M\to\infty}
        b-a,\qquad 0\le a<b\le 1\;,$$
where $\{t\}$ is the fractional part of a real number $t$.
In terms of Definition 1, $(x_n)$ is uniformly distributed modulo~1 if and
only if the sequence $(\{x_n\})$ of fractional parts is $F$-distributed, where
$F$ is the distribution function of the uniform distribution on $[0,1]$:
$$F(x)=\cases{
                        0,&$x<0,$\cr
                        x,&$0\le x\le 1,$\cr
                        1,&$x>1.$\cr
              } $$
The generalization of the notion of an asymptotically $F$-distributed sequence
to that of an asymptotically well $F$-distributed sequence clearly carries over
to our case. Instead of requiring only that the dispersion of large initial
pieces of the sequence becomes more and more even, we require this to
happen at arbitrary locations. This version is termed
{\it well-distribution}. Thus, $(x_n)_{n=1}^\infty$ is
{\it well-distributed modulo}~$1$ if
$$\frac{\left|\{L<n\le M:a\le \{x_n\}<b\}\right|}{M-L}\longto_{M-L\to\infty}
        b-a,\qquad 0\le a<b\le 1\;.$$
Both notions have multi-dimensional analogue. A sequence $(\x_n)_{n=1}^\infty$ in
$\R^s$ is {\it uniformly distributed modulo~1 in} $\R^s$ if
$$\frac{\left|\{1\le n\le N:\bfa\le \{\bfx_n\}<\bfb\}\right|}{N}\longto_{N\to\infty}
        \prod_{i=1}^s (b_i-a_i),\qquad \0\le \bfa<\bfb \le \1\;,$$
where inequalities between vectors in $\R^s$ are to be understood component-wise,
$\0=(0,0,\ldots,0)\in\R^s,\ \bfa=(a_1,a_2,\ldots,a_s)$, and so forth.

Perhaps the most basic example of a sequence which is uniformly distributed modulo~1
is $(n\alpha)_{n=1}^\infty$, where $\alpha$ is an arbitrary irrational. In the
multi-dimensional case, the sequence $(n\alpha_1,n\alpha_2,\ldots,n\alpha_s)$
is uniformly distributed modulo~1 in $\R^s$ if and only if the numbers
$1,\alpha_1,\alpha_2,\ldots,\alpha_s$ are linearly independent over $\Q$.
Moreover, in this case uniform distribution implies well-distribution
(cf. \cite[Example 1.6.1, Exercise 1.6.14]{KN}).

Given a partition $\N=\bigcup_{j=1}^l A_j$ and positive integers $r_j,\;j=1,\ldots,l$,
we define the $(r_j)_{j=1}^l$-{\it inflation} of the given partition as the partition
of $\N$ obtained by inflating each element of each of the sets $A_j$ into $r_j$
elements. More precisely, we construct sets $B_j,\;j=1,\ldots,l$, as follows.
For a positive integer $i$, let $f(i)=j$ if $i\in A_j$. Given
any positive integer $n$, let $m$ be defined by
$\sum_{i=1}^{m-1} f(i)< m\le \sum_{i=1}^{m} f(i)\;.$ Let $n\in B_j$ if $m\in A_j$.
The following lemma is routine.

\proclaim Lemma 2. In this setup:
\ben
\item
If $\D(A_j)=d_j,\ 1\le j\le l$, then $\D(B_j)=\frac{r_j d_j}{\sum_{i=1}^l r_i d_i}\;.$
\item
If $\BD(A_j)=d_j,\ 1\le j\le l$, then $\BD(B_j)=\frac{r_j d_j}{\sum_{i=1}^l r_i d_i}\;.$
\een

\bdemo Proof of Theorem 1. Between any two consecutive odd multiples of $w_0$, there
is at most one odd multiple of each $w_j,\ 1\le j\le K-1$. In fact, one easily verifies
that, given a positive integer $m$, there is an odd multiple of $w_j$ between
$(2m-1)w_0$ and $(2m+1)w_0$, namely there exists an integer $n$ with
\begin{eqnarray}
(2m-1)w_0\le (2n-1)w_j<(2m+1)w_0,\label{contain_w_j}
\end{eqnarray}
if and only if
\begin{eqnarray}
m c_j\in\left({1-c_j\over 2},{1+c_j\over 2}\right]\;(\mod 1)\;.\label{cond_contain_w_j}
\end{eqnarray}
Moreover, the relative position of $(2n-1)w_j$ within the interval
$[(2m-1)w_0,(2m+1)w_0)$ is the same, but in the opposite direction, as that of
$m c_j(\mod 1)$ within the interval $\big({1-c_j\over 2},{1+c_j\over 2}\big]$, that is
\begin{eqnarray}
(2n-1)w_j=\alpha\cdot(2m-1)w_0+(1-\alpha)\cdot(2m+1)w_0,\qquad (0<\alpha\le 1)\;,
\label{position_w_j}
\end{eqnarray}
if and only if
\begin{eqnarray}
m c_j\equiv (1-\alpha)\cdot{1-c_j\over 2}+\alpha\cdot{1+c_j\over 2}\;(\mod 1)\;.
\label{cond_position_w_j}
\end{eqnarray}

Next we define a partition of $\N$ as follows. Write the elements of $\Delta$
in ascending order:
$\Delta=\{\delta_1<\delta_2<\delta_3<\ldots\}$. Given $n\in\N$, let
$S\subseteq\{1,2,\ldots,K-1\}$ denote the set of all those $j$'s such that the unique
interval of the form $[(2m-1)w_0,(2m+1)w_0)$ containing $\delta_n$ contains an odd
multiple of $w_j$. The set of all integers $n$ giving rise in this way to any set $S$
is denoted by $B_S$. Consider the partition
$\N=\bigcup_{S\subseteq\{1,2,\ldots,K-1\}} B_S$.
To prove the theorem using Lemma~1, we have to find the Banach densities of the sets
$B_S$ and the asymptotic distribution of the corresponding subsequences
$(\chi_n)_{n\in B_S}$ of $\chi_n$.

The partition of $\N$ into sets of the form $B_S$ is obtained as an inflation of a
somewhat more straightforward partition. In fact, let $S$ be any subset of
$\{1,2,\ldots,K-1\}$. Denote by $A_S$ the set of those positive integers $n$ for which
the interval $[(2n-1)w_0,(2n+1)w_0)$ contains odd multiples of $w_j$ for $j\in S$ and
does not contain such multiples of the other $w_j$'s. Then
$\N=\bigcup_{S\subseteq\{1,\ldots,K-1\}} A_S$ is a partition, and its
$(|S|+1)_{S\subseteq\{1,2,\ldots,K-1\}}$-inflation yields the partition
$\N=\bigcup_{S\subseteq\{1,2,\ldots,K-1\}} B_S$.

In view of the equivalence of (\ref{contain_w_j}) and (\ref{cond_contain_w_j}), $A_S$
is the set of those $n$'s for which $n c_j\in\left({1-c_j\over 2},{1+c_j\over 2}\right]$
for $j\in S$ and $n c_j\notin\left({1-c_j\over 2},{1+c_j\over 2}\right]$ for $j\notin S$.
By the conditions of the theorem, the numbers $1,c_1,\ldots,c_{K-1}$ are linearly
independent over $\Q$, and hence the sequence
$\c=(nc_1,nc_2,\ldots,nc_{K-1})_{n=1}^\infty$
is well-distributed modulo~1 in $\R^{K-1}$. This means that
\begin{eqnarray}
\D(A_S)=\BD(A_S)=\prod_{i\in S} c_i \prod_{i\notin S} (1-c_i)\;.\label{density_A_S}
\end{eqnarray}
Denote the right hand side of (\ref{density_A_S}) by $p_S$.
In view of the above and Lemma 2, this implies
\begin{eqnarray}
\D(B_S)=\BD(B_S)=\frac{(|S|+1) p_S}{\sum_{T\subseteq\{1,2,\ldots,K-1\}} (|T|+1) p_T}\;.
\label{density_B_S}
\end{eqnarray}
The denominator on the right hand side can be given a simpler form. In fact, let
$X_i,\ i=1,2,\ldots,K-1,$ be independent random variables with $X_i\sim B(1,c_i)$,
and $X=\sum_{i=1}^{K-1} X_i$. Then:
\begin{eqnarray}
\sum_{T\subseteq\{1,2,\ldots,K-1\}} (|T|+1) p_T=E(X+1)=1+c_1+\ldots+c_{K-1}=s\;.
\end{eqnarray}
Hence:
\begin{eqnarray}
\BD(B_S)=\frac{(|S|+1) p_S}{s}\;.
\label{density_B_S_2nd_form}
\end{eqnarray}

Let $S$ be an arbitrary fixed subset of $\{1,2,\ldots,K-1\}$, say $S=\{1,2,\ldots,l\}$,
where $0\le l\le K-1$. If $n\in A_S$, then there exist odd integers
$a_{1n},a_{2n},\ldots,a_{ln}$ such that $a_{jn}w_j\in [(2n-1)w_0,(2n+1)w_0)$. Put:
$$v_n=(a_{1n}w_1,a_{2n}w_2,\ldots,a_{ln}w_l)-(2n-1)w_0\cdot(1,1,\ldots,1)
  \in [0,2w_0)^l,\qquad n\in A_S\;.$$
By the equivalence of (\ref{position_w_j}) and (\ref{cond_position_w_j}), the sequence
$(v_n)_{n\in A_S}$ is well-distributed modulo~$2w_0$ in~$\R^l$. Now each $v_n$ gives
rise to $l+1$ terms of $(\chi_n)_{n\in B_S}$, as follows. Let
$v_n^{(1)}\le v_n^{(2)}\le\ldots\le v_n^{(l)}$ be all coordinates of $v_n$ in
ascending order. Set:
$$u_n=(v_n^{(1)},v_n^{(2)}-v_n^{(1)},\ldots,v_n^{(l)}-v_n^{(l-1)},2w_0-v_n^{(l)}),
  \qquad n\in A_S\;.$$
The sequence $(\chi_n)_{n\in B_S}$ consists of of all coordinates of all vectors
$u_n$. Now we use the fact that if $X_1,X_2,\ldots,X_r$ are independent random variables,
distributed ${\rm U}(0,h)$, and $X^{(1)},X^{(2)},\ldots,X^{(r)}$ are the corresponding
order statistics, then each of the random variables
$X^{(1)},X^{(2)}-X^{(1)},\ldots,X^{(r)}-X^{(r-1)},h-X^{(r)}$ has the distribution
function defined by
$G(x)=1-(x/h)^r$ for $0\le x\le h$ (which follows as a special case from
\cite[p.42, ex.23]{Fe}). Consequently, for each $1\le j\le l+1$, the sequence given
by the $j$th coordinate of all vectors $u_n,\ n\in A_S$, is asymptotically well
$G_1$-distributed, where $G_1(x)=1-(x/2w_0)^l$ for $0\le x\le 2w_0$. Hence the
sequence $(\chi_n)_{n\in B_S}$ is asymptotically well $G_1$-distributed. Combined
with (\ref{density_B_S_2nd_form}), it proves (\ref{eq_P_2(x)}).

We shall indicate only briefly the proof of (\ref{eq_P_1(x)}), which is quite
simpler. This time, we split $(\chi_n)$ into a union of subsequences
$(\chi_n^{(i)}),\ 0\le i\le K-1$, by putting $\chi_n$ in the sequence
$\chi_n^{(i)}$ if $\delta_n$ is a multiple of $w_i$. Clearly, the proportion
of terms of $(\chi_n)$ belonging to $(\chi_n^{(i)})$ is $c_i/s$. Next, consider
the minimal odd multiples of all $w_j$'s which are larger than $\delta_n$. The
minimum of these $K$ numbers is $\delta_{n+1}$. For each $j\ne i$, the distance from
$\delta_n$ to the minimal odd multiple of $w_j$ following $\delta_n$ is ``distributed"
${\rm U}(0,2w_j)$. (For $i=0$ it is also possible that the next term will be again a
multiple of $w_0$.) The linear independence of the $C_i$'s over $\Q$ implies that these
$K-1$ distances are (statistically) independent, so that their minimum is distributed
according to the function
$G_2(x)=1-\prod_{j=0\atop j\ne i}^{K-1}\left(1-\frac{x}{2w_j}\right)$
on the interval $[0,2w_0)$. These considerations can be formalized to prove
(\ref{eq_P_1(x)}). This completes the proof.
\edemo

\demo Remark. It is possible to shorten the proof by proving directly the equality
of the right hand sides of (\ref{eq_P_1(x)}) and (\ref{eq_P_2(x)}).
In fact, it is easy to integrate both forms with respect to $x$; the
equality of the resulting expressions follows easily from the binomial theorem.
We have chosen the long way, as it is more instructive.

\bdemo Proof of Theorem 2. The distribution $F_K$ is obtained from that in Theorem~1
by stretching by the constant factor $\frac{s}{2w_0}$. Hence:
\begin{eqnarray}
F_K(x)&=&\cases{
               0,&$\ x<0,$\cr
               \displaystyle{1-\frac{1}{s}\sum_{i=0}^{K-1} c_i\prod_{j=0\atop j\ne i}^{K-1}
                   \left(1-\frac{c_j x}{s}\right),}&$\ 0\le x<s,$\cr
               1,&$\ s\le x\;.$\cr
              }
\label{eq_F_K(x)}
\end{eqnarray}
Note that some of the values appearing on the right hand side depend on $K$ implicitly.
Namely, since $w_0$ is assumed in Theorem~1 to be the least $w_i$, each time a $C_i$
is selected which is larger than all the heretofore selected $C_j$'s, we have to
rearrange the $C_j$'s, thus changing $w_0$ and the $c_j$'s. We have to show that
\begin{eqnarray}
F_K(x)\longto_{K\to\infty} 1-e^{-x},\qquad x\ge 0\;.
\label{conv_F_K(x)}
\end{eqnarray}
Indeed, fix $x\ge 0$. Since
\begin{eqnarray}
s=c_0+c_1+\ldots+c_{K-1}=\frac{C_0+C_1+\ldots+C_{K-1}}{C_0}\ge C_0+C_1+\ldots+C_{K-1}
\end{eqnarray}
and the $C_i$'s are independent and uniformly distributed in $[0,1]$, we have
\begin{eqnarray}
s\longto_{K\to\infty}^{\rm a.s.}\infty\;.
\label{conv_s}
\end{eqnarray}
Hence, with probability 1, for sufficiently large $K$ we have
\begin{eqnarray}
F_K(x)=\displaystyle{1-\frac{1}{s}\sum_{i=0}^{K-1} c_i\prod_{j=0\atop j\ne i}^{K-1}
                   \left(1-\frac{x}{2w_j}\right)}\;.
\end{eqnarray}
Thus, to prove (\ref{conv_F_K(x)}) we need to show that
\begin{eqnarray}
\frac{1}{s}\sum_{i=0}^{K-1} c_i\prod_{j=0\atop j\ne i}^{K-1}
\left(1-\frac{c_j x}{s}\right) \longto_{K\to\infty}^{\rm a.s.} e^{-x},\qquad x\ge 0\;.
\end{eqnarray}
Now, on the one hand, using the inequality
$$1-t\le e^{-t},\qquad t\in\R\;,$$
we have
$$\prod_{j=0\atop j\ne i}^{K-1}\left(1-\frac{c_j x}{s}\right)\le
e^{-x\sum_{j=0\atop j\ne i}^{K-1}\frac{c_j}{s}}\le
e^{-x+x/s},\qquad i=0,1,\ldots,K-1\;,$$
and therefore
\begin{eqnarray}
\frac{1}{s}\sum_{i=0}^{K-1} c_i\prod_{j=0\atop j\ne i}^{K-1}
\left(1-\frac{c_j x}{s}\right) \le \frac{1}{s}\sum_{i=0}^{K-1} c_i e^{-x+x/s}
=e^{-x+x/s}\longto_{K\to\infty}^{\rm a.s.} e^{-x}\;.
\label{upper_bound}
\end{eqnarray}
On the other hand, as $t\to 0$ we have
$$e^{-(t+t^2)}=1-(t+t^2)+\frac{(t+t^2)^2}{2}+O(t^3)=1-t-\frac{t^2}{2}+O(t^3)\;,$$
so that for all $t$ in some sufficiently small neighborhood of $0$
$$e^{-(t+t^2)}\le 1-t\;.$$
Consequently:
\begin{eqnarray}
\prod_{j=0\atop j\ne i}^{K-1}\left(1-\frac{c_j x}{s}\right)\ge
e^{-x\sum_{j=0\atop j\ne i}^{K-1}\frac{c_j}{s}
   -x^2\sum_{j=0\atop j\ne i}^{K-1}\frac{c_j^2}{s^2}}\ge e^{-x-K x^2/s^2}\;.
\label{lower_bound_1}
\end{eqnarray}
Obviously, with probability 1, $s$ grows linearly with $K$, namely for
all sufficiently large $K$ we have $s\ge aK$ for a suitably chosen $a>0$.
(In fact, any $a<\frac{1}{2}$ will do.) By (\ref{lower_bound_1}):
\begin{eqnarray}
\prod_{j=0\atop j\ne i}^{K-1}\left(1-\frac{c_j x}{s}\right)\ge
e^{-x-K x^2/s^2}\longto_{K\to\infty}^{\rm a.s.} e^{-x}\;.
\label{lower_bound_2}
\end{eqnarray}
From (\ref{upper_bound}) and (\ref{lower_bound_2}) it follows that
\begin{eqnarray}
\frac{1}{s}\sum_{i=0}^{K-1} c_i\prod_{j=0\atop j\ne i}^{K-1}
\left(1-\frac{c_j x}{s}\right)
\longto_{K\to\infty}^{\rm a.s.} e^{-x}\;,
\end{eqnarray}
which completes the proof.
\edemo

\begin{acknowledgments}
This work was supported in part by grants from the Israel Academy of 
Science and Humanities under the program {\em Centers for Excellence}, by the Basic
Research Foundation and the BS -- Binational Israel-US Foundation.
\end{acknowledgments}


\end{document}